\documentclass[preprint,prd,aps,nofootinbib]{revtex4}
\usepackage{amsmath}
\usepackage{graphics}

\begin{document}

\def\cstok#1{\leavevmode\thinspace\hbox{\vrule\vtop{\vbox{\hrule\kern1pt
\hbox{\vphantom{\tt/}\thinspace{\tt#1}\thinspace}}
\kern1pt\hrule}\vrule}\thinspace}

\begin{center}
\bibliographystyle{article}
{\Large \textsc{Gravitational waves about curved backgrounds:
a consistency analysis in de Sitter spacetime}}
\end{center}

\author{Donato Bini,$^{1,2}$ \thanks{
Electronic address: binid@icra.it} Salvatore Capozziello,$^{3,4}$
\thanks{ Electronic address: capozziello@na.infn.it} Giampiero
Esposito$^{4}$ \thanks{ Electronic address:
giampiero.esposito@na.infn.it}}

\affiliation{
${\ }^{1}$CNR, Istituto per le Applicazioni del Calcolo ``M. Picone'',
Via del Policlinico 137, 00161 Rome, Italy\\
${\ }^{2}$International Center for Relativistic Astrophysics - I.C.R.A.,
University of Rome ``La Sapienza,'' 00185 Rome, Italy\\
${\ }^{3}$Dipartimento di Scienze Fisiche, Complesso Universitario di Monte
S. Angelo,\\
Via Cintia, Edificio 6, 80126 Napoli, Italy\\
${\ }^{4}$Istituto Nazionale di Fisica Nucleare, Sezione di Napoli,\\
Complesso Universitario di Monte S. Angelo, Via Cintia, Edificio 6, 80126
Napoli, Italy}

\vspace{0.4cm}
\date{\today}

\begin{abstract}
Gravitational waves are considered as metric
perturbations about a curved background metric, rather than the
flat Minkowski metric since several situations of physical
interest can be discussed by this generalization. In this case,
when the de Donder gauge is imposed, its preservation under
infinitesimal spacetime diffeomorphisms is guaranteed if and only
if the associated covector is ruled by a second-order hyperbolic
operator which is the classical counterpart of the ghost operator
in quantum gravity. In such a wave equation, the Ricci term has
opposite sign with respect to the wave equation for Maxwell theory
in the Lorenz gauge. We are, nevertheless, able to relate the
solutions of the two problems, and the algorithm is applied to the
case when the curved background geometry is the de Sitter
spacetime. Such vector wave equations are studied in two different
ways: $i)$ an integral representation, $ii)$ through a solution
by factorization of the hyperbolic equation. The latter method is
extended to the wave equation of metric perturbations in the de
Sitter spacetime. This approach is a step towards a general
discussion of gravitational waves in the de Sitter spacetime and
might assume relevance in cosmology in order to study the
stochastic background emerging from inflation.
\end{abstract}

\maketitle
\bigskip
\vspace{2cm}

\section{Introduction}

One of the longstanding problems of modern gravitational physics
is the detection of gravitational waves, for which the standard
theoretical analysis relies upon the split of the space-time
metric $g_{ab}$ into ``background plus perturbations'', i.e.
\begin{equation}
g_{ab}=\gamma_{ab}+h_{ab},
\label{(1)}
\end{equation}
where $\gamma_{ab}$ is the background Lorentzian metric, often
taken to be of the Minkowski form $\eta_{ab}$, while the symmetric
tensor field $h_{ab}$ describes perturbations about $\gamma_{ab}$.
However, the background $\gamma_{ab}$ needs not to be Minkowskian
in several cases of physical interest, nor it has  to be always a
solution of the vacuum Einstein equations. As a consequence, we
are therefore aiming to investigate in more detail what happens if
the background space-time $(M,\gamma_{ab})$ has a non-vanishing
Riemann curvature.

This issue has to be seriously considered from an experimental
point of view since the gravitational wave detectors of new
generation are designed also to investigate strong-field regimes:
this means that the physical situations, where only  the standard
Minkowski background is taken into account, could be misleading
in order to achieve self-consistent results.

In particular, several ground-based laser interferometers have
been built in the United States (LIGO) \cite{ligo}, Europe (VIRGO
and GEO) \cite{virgo,geo}, and Japan (TAMA) \cite{tama} and are
now in the data taking phase for frequency ranges about
$10^{-1}kHz$. However, new advanced optical configurations allow
to reach sensitivities slightly above and below the  standard
quantum limit for free test-particles, hence we are now approaching
the epoch of second \cite{buonanno1} and third \cite{buonanno2}
generation of gravitational wave detectors. This fact, in
principle, allows to investigate wide ranges of frequencies where
strong field regimes or alternative theories of gravity can be
considered \cite{extended1,extended2,extended3}. Besides, the
laser interferometer space antenna (LISA) \cite{lisa} (which is
mainly devoted to work in the range $10^{-4}\sim 10^{-2} Hz$)
should fly within the next decade principally aimed at investigating
the stochastic background of gravitational waves. At much lower
frequencies $(10^{-17}Hz)$, cosmic microwave background (CMB)
probes, like the forthcoming PLANCK satellite, are designed to
detect also gravitational waves by measuring the CMB polarization
\cite{planck} while millisecond pulsar timing can set interesting
upper limits in the frequency range between $10^{-9}\sim
10^{-8}Hz$ \cite{jenet}. At these frequencies, the large number of
millisecond pulsars detectable by the square kilometer array would
provide a natural ensemble of
clocks which can be used as multiple arms of a gravitational wave
detector \cite{skate}.

This forthcoming experimental situation is intriguing but deserves
a serious theoretical analysis which cannot leave aside the
rigorous investigation of strong-field regimes and the possibility
that further polarization states of gravitational waves could come
out in such regimes. For example, if one takes into account
scalar-tensor theories of gravity \cite{extended1} or higher-order
theories \cite{extended2}, scalar-massive gravitons should be
considered. This implies that the standard approach where 
gravitational waves are assumed as small perturbations (coming
only from Einstein's general relativity) on a Minkowski
background could be totally insufficient. On the other hand,  the
existence of these further polarization modes could be a
straightforward solution of the dark matter problem since massive
gravitons could be testable cold dark matter candidates as
discussed in \cite{depaula,dubovsky}.

In this paper, we want to face the issue of the rigorous
formulation of gravitational wave problem in curved backgrounds.
In particular, we want to perform a consistency analysis of
gravitational waves in the de Sitter spacetime. Achieving
solutions in this maximally symmetric background could constitute
the paradigm to investigate any curved spacetime by the same
techniques and could have interesting cosmological applications if
a conformal analysis is undertaken as, for example in
\cite{extended3}, where it is shown how the amplitude of
cosmological gravitational waves strictly depends on the
cosmological background.

It is straightforward to show that, in a covariant formulation,
the supplementary condition for gravitational waves can be
described by a functional $\Phi_{a}$ acting on the space of
symmetric rank-two tensors $h_{ab}$ occurring in Eq. (1). For any
choice of $\Phi_{a}$, one gets a different realization of the
invertible operator $P_{ab}^{\; \; \; cd}$ on metric
perturbations. The basic equations of the theory read therefore as
\begin{equation}
P_{ab}^{\; \; \; cd}h_{cd}=0,
\label{(2)}
\end{equation}
\begin{equation}
\Phi_{a}(h)=0,
\label{(3)}
\end{equation}
where $P_{ab}^{\; \; \; cd}$ results from the expansion of the
action functional to quadratic order in the metric perturbations.
A deep link exists between classical and quantum theory, since in
the latter, the one-loop analysis depends on the functional
determinant of $P_{ab}^{\; \; \; cd}$, after requiring that all
metrics in Eq. (1) are positive-definite, i.e. Riemannian. Our
analysis will instead be Lorentzian and classical.

The layout of the paper is the following. Section 2 studies the de
Donder choice for $\Phi_{a}$ and its preservation, while Secs. 3
and 4 deal with massless Green functions in de Sitter spacetime.
This is done because the problem of preserving Eq. (3) under
infinitesimal diffeomorphisms leads precisely to vector wave
equations. These are solved by an integral representation or by
separation of variables. This analysis prepares the ground for
studying the wave equation on metric perturbations itself through
separation of variables, in Sec. 6. Concluding remarks and open
problems are presented in Sec. 7, while relevant details are
given in the Appendix.

\section{Preservation of the de Donder supplementary condition}

Our first concern is how to implement in a consistent way the choice of
supplementary condition. In general relativity, this is taken to be
of the de Donder type (below $h \equiv \gamma^{cd}h_{cd}$)
\begin{equation}
\Phi_{a}(h)=\nabla^{b}\left(h_{ab}-\frac{1}{2}\gamma_{ab}h
\right), \label{(4)}
\end{equation}
if one wants to obtain the standard covariant wave operator on metric
perturbations, where $\nabla^{b}$ denotes covariant derivative with respect
to the background metric $\gamma_{ab}$. Under infinitesimal space-time
diffeomorphisms, the metric perturbations suffer the variation (the
round brackets denoting symmetrization)
\begin{equation}
\delta h_{ab}=\nabla_{(a} \; \varphi_{b)},
\label{(5)}
\end{equation}
where $\varphi_{b}$ is a covector, with associated one-form
$\varphi_{b}dx^{b}$ and vector field $\varphi^{a}\frac{\partial}
{\partial x^{a}}$ (having set $\varphi^{a} \equiv
\gamma^{ab}\varphi_{b}$, which results from the isomorphism
between tangent and cotangent space to the background space-time,
that turns covectors into vectors, or the other way around). The
change suffered from the de Donder gauge in (4) when metric
perturbations are varied according to (5) is then found to be
\begin{equation}
\delta \Phi_{a}(h)=-\left(\delta_{a}^{\; b}\cstok{\ }
+R_{a}^{\; b} \right)\varphi_{b},
\label{(6)}
\end{equation}
where $\cstok{\ }$ is the standard d'Alembert operator in curved
space-time, i.e. $\cstok{\ } \equiv
\gamma^{cd}\nabla_{c}\nabla_{d}$. By virtue of Eqs. (4) and (6), if
the de Donder gauge was originally satisfied, it is preserved
under space-time diffeomorphisms if and only if $\varphi_{b}$
solves the equation $\delta \Phi_{a}(h)=0$. At this stage, to
fully exploit what is known about the wave equation for Maxwell
theory in curved space-time in the Lorenz gauge\footnote{In
\cite{Lore67}, the author L. Lorenz, who was studying the identity
of the vibrations of light with electrical currents, built a set
of retarded potential for electrodynamics which, with hindsight,
can be said to satisfy the gauge condition
$\nabla^{\mu}A_{\mu}=0$, which therefore should not be ascribed to
H. Lorentz.} we bear in mind that this reads as
\begin{equation}
\left(-\delta_{a}^{\; b}\cstok{\ }+R_{a}^{\; b}\right)A_{b}=0.
\label{(7)}
\end{equation}
This suggests adding $R_{a}^{\; b}\; \varphi_{b}$ to both sides of
$\delta \Phi_{a}(h)=0$ (see (6)),
so as to cast it eventually in the form
\begin{equation}
P_{a}^{\; b} \; \varphi_{b}=2R_{a}^{\; b} \; \varphi_{b},
\label{(8)}
\end{equation}
where
\begin{equation}
P_{a}^{\; b} \equiv -\delta_{a}^{\; b}\cstok{\ }+R_{a}^{\; b}
\label{(9)}
\end{equation}
is the standard gauge-field operator (see round brackets in
Eq. (7)) in the Lorenz gauge. For this operator, the inverse
${\widetilde P}_{a}^{\; b}$ is an integral operator with kernel
given by the photon Green function, so that we can solve Eq. (8) in
the form
\begin{equation}
\varphi_{c}=\varphi_{c}^{(0)}+2{\widetilde P}_{c}^{\; a} \;
R_{a}^{\; b} \; \varphi_{b},
\label{(10)}
\end{equation}
where $\varphi_{c}^{(0)}$ is a solution of the homogeneous wave
equation \cite{Frie75}
\begin{equation}
P_{a}^{\; b} \; \varphi_{b}^{(0)}=0,
\label{(11)}
\end{equation}
while ${\widetilde P}_{c}^{\; a}$ is the inverse operator, satisfying
\begin{equation}
{\widetilde P}_{c}^{\; a} \; P_{a}^{\; b}=\delta_{c}^{\; b}.
\label{(12)}
\end{equation}
This ${\widetilde P}_{c}^{\; a}$ is an integral operator with kernel
given by the massless spin-1 Green function $G_{ab}(x,x') \equiv G_{ab'}$.
The latter can be chosen, for example,
to be of the Feynman type, i.e. that solution
of the equation (see Appendix for the notation)
\begin{equation}
\left(-\delta_{a}^{\; b} \cstok{\ }+R_{a}^{\; b} \right)
G_{bc'}=g_{ac'}{\delta(x,x') \over \sqrt{-\gamma}},
\label{(13)}
\end{equation}
having the asymptotic expansion as $\sigma \rightarrow 0$
\cite{APNYA-9-220,Bimo04}
\begin{equation}
G_{ab'} \sim {{\rm i}\over 8 \pi^{2}}\left[\sqrt{\bigtriangleup}
{g_{ab'}\over (\sigma+{\rm i}\varepsilon)}
+V_{ab'}\log (\sigma+{\rm i}\varepsilon)+W_{ab'} \right],
\label{(14)}
\end{equation}
where $\sigma(x,x')$ is the Ruse--Synge world
function \cite{Ruse31, Syng31, Syng60}, equal to half the
square of the geodesic distance $\mu$ between the points $x$ and $x'$.

\section{Massless Green functions in de Sitter spacetime}

This general scheme can be completely implemented in the
relevant case \cite{Hawk00} of de Sitter space where, relying upon
the work in \cite{Alle86}, we know that the massless spin-1 Green
function reads as
\begin{equation}
G_{ab'}=\alpha(\mu)g_{ab'}+\beta(\mu)n_{a}n_{b'},
\label{(15)}
\end{equation}
where $\mu(x,x') \equiv \sqrt{2\sigma(x,x')}$ is the geodesic distance
between $x$ and $x'$, $n^{a}(x,x')$ and $n^{a'}(x,x')$ are the unit
tangents to the geodesic at $x$ and $x'$, respectively, for which
\begin{equation}
n_{a}(x,x')=\nabla_{a} \mu(x,x'), \;
n_{a'}(x,x') = \nabla_{a'} \mu(x,x'),
\label{(16)}
\end{equation}
while, in terms of the variable
\begin{equation}
z \equiv {1\over 2}\left(1+\cos {\mu \over \rho} \right),
\label{(17)}
\end{equation}
the coefficient functions $\alpha$ and $\beta$ are given, in four
dimensions, by \cite{Alle86}
\begin{equation}
\alpha(z)={1\over 48 \pi^{2} \rho^{2}}\left[{3\over (1-z)}
+{1\over z}+\left({2\over z}+{1\over z^{2}}\right)\log(1-z) \right],
\label{(18)}
\end{equation}
\begin{equation}
\beta(z)={1\over 24 \pi^{2}\rho^{2}}\left[1-{1\over z}
+\left({1\over z}-{1\over z^{2}}\right)\log(1-z) \right].
\label{(19)}
\end{equation}
Strictly speaking, the formulae (18)--(19) are first derived in the
Euclidean de Sitter space. In the Lorentzian de Sitter
spacetime $M$ which is what we are interested in, one can define the set
\cite{Alle86}
\begin{equation}
J_{x} \equiv \left \{ x' \in M: \exists \; {\rm geodesic} \;
{\rm from} \; x \; {\rm to} \; x' \right \}.
\label{(20)}
\end{equation}
Moreover, it is well-known that $M$ can be viewed as an hyperboloid
imbedded in flat space, i.e. as the set of points $Y^{a} \in {\bf R}^{n+1}$
such that
\begin{equation}
Y^{a}Y^{b}\eta_{ab}=\rho^{2},
\label{(21)}
\end{equation}
where $\eta_{ab}={\rm diag}(-1,1,...,1)$, so that its induced metric
reads as
\begin{equation}
ds^{2}=\eta_{ab}dY^{a}dY^{b}.
\label{(22)}
\end{equation}
As is stressed in Ref. \cite{Alle86}, the relation
\begin{equation}
z(x,x')={1\over 2} \left[1+{\eta_{ab}Y^{a}(x)Y^{b}(x')\over \rho^{2}}
\right]
\label{(23)}
\end{equation}
is well defined both inside and outside $J_{x}$, and it is an analytic
function of the coordinates $Y^{a}$. Thus, Eq. (23) makes it possible to
define $z(x,x')$ everywhere on de Sitter,
and one can define the geodesic distance
\begin{equation}
\mu(x,x') \equiv 2\rho \cos^{-1}(\sqrt{z})
\label{(24)}
\end{equation}
as the limiting value \cite{Alle86} above the standard branch cut of
$\cos^{-1}$. Along similar lines, the equations defining $n_{a},n_{a'}$
and $g_{ab'}$ have right-hand sides which are analytic functions of the
coordinates $Y^{a}$, and are hence well defined everywhere on Lorentzian
de Sitter spacetime \cite{Alle86}.

\section{Evaluation of the kernel}

In a de Sitter background the Ricci tensor is proportional to the metric
through the cosmological constant: $R_{ab}=\Lambda g_{ab}$, and hence the
formulae (10), (15), (18) and (19) lead to the following explicit
expression for the solution of the inhomogeneous wave equation (8):
\begin{equation}
\varphi_{c}(x)=\varphi_{c}^{(0)}(x)+2 \Lambda \int
\Bigr[\alpha(z(\mu(x,x')))g_{c}^{\; a'}
+\beta(z(\mu(x,x')))n_{c}n^{a'}\Bigr]\varphi_{a'}(x')
\sqrt{-\gamma(x')}d^{4}x',
\label{(25)}
\end{equation}
where, from Eq. (24),
\begin{equation}
\mu(x,x')=2\rho \cos^{-1} \sqrt{{1\over 2}\left(
1+{\eta_{ab}Y^{a}(x)Y^{b}(x')\over \rho^{2}}\right)},
\label{(26)}
\end{equation}
while Eqs. (18) and (19) should be exploited to express $\alpha$ and
$\beta$, bearing in mind Eq. (26) jointly with
\begin{equation}
z(x,x')={1\over 2}\left[1+\cos \left({\mu(x,x')\over \rho}\right)\right].
\label{(27)}
\end{equation}
Moreover, the bivector $g_{c}^{\; a'}$ in the integrand (25) is given
by \cite{Alle86}
\begin{equation}
g_{a}^{\; b'}=C^{-1}(\mu)\nabla_{a}n^{b'}-n_{a}n^{b'}, \;
C(\mu)=-{1\over \rho \sin (\mu / \rho)}.
\label{(28)}
\end{equation}
The right-hand side of the formula expressing $g_{a}^{\; b'}$ is an
analytic function of the coordinates $Y^{a}$ and is therefore well
defined everywhere on de Sitter \cite{Alle86}. The
integral on the right-hand side of Eq. (25) can be conveniently expressed
the form
\begin{equation}
f_{c}(x)=\int \Bigr[\alpha(z)C^{-1}(\mu)\nabla_{c}\nabla^{a'}\mu
+(\beta(z)-\alpha(z))(\nabla_{c}\mu)(\nabla^{a'}\mu)\Bigr]
\varphi_{a'}(x')\sqrt{-\gamma(x')}d^{4}x',
\label{(29)}
\end{equation}
with $\alpha$ and $\beta-\alpha$ given by (cf. (18) and (19))
\begin{equation}
\alpha(z)={(1+2z)\over 48 \pi^{2}\rho^{2}}\left[{1\over z(1-z)}
+{1\over z^{2}}\log(1-z)\right],
\label{(30)}
\end{equation}
\begin{equation}
\beta(z)-\alpha(z)={1\over 48 \pi^{2}\rho^{2}}\left[
{(-3+2z-2z^{2})\over z(1-z)}-{3\over z^{2}}\log(1-z)\right].
\label{(31)}
\end{equation}
Equation (25) is therefore an integral equation reading as
\begin{equation}
\varphi_{c}(x)=\varphi_{c}^{(0)}(x)+\Lambda \int
K_{c}^{\; a'}\varphi_{a'} \sqrt{-\gamma(x')}d^{4}x',
\label{(32)}
\end{equation}
with unbounded kernel given by
\begin{equation}
K_{c}^{\; a'} \equiv 2 \Bigr[\alpha(z)C^{-1}(\mu)
\nabla_{c}\nabla^{a'}\mu
+(\beta(z)-\alpha(z))(\nabla_{c}\mu)(\nabla^{a'}\mu)\Bigr].
\label{(33)}
\end{equation}
This kernel is indeed unbounded by virtue of the limits
\begin{equation}
48 \pi^{2} \rho^{2} \lim_{z \to 0} z \alpha(z)={1\over 2},
\label{(34)}
\end{equation}
\begin{equation}
48 \pi^{2} \rho^{2} \lim_{z \to 1} (1-z) \alpha(z)=1,
\label{(35)}
\end{equation}
\begin{equation}
48 \pi^{2} \rho^{2} \lim_{z \to 0} z (\beta(z)-\alpha(z))
=-{3\over 2},
\label{(36)}
\end{equation}
\begin{equation}
48 \pi^{2} \rho^{2} \lim_{z \to 1} (1-z)(\beta(z)-\alpha(z))=-3.
\label{(37)}
\end{equation}
At this stage, we can exploit (23) and (33) to re-express the kernel in
the form
\begin{eqnarray}
K_{c}^{\; a'}&=& {(\nabla_{c}z)(\nabla^{a'}z)
\over 24 \pi^{2}\rho^{4}(1-z)}\biggr[2+
\left(-3 +{\sqrt{z}\over 2}(1+2z)\right)
\left({1\over z(1-z)}+{1\over z^{2}}\log(1-z) \right)\biggr]
\nonumber \\
&+& {(\nabla_{c}\nabla^{a'}z)\over 6 \pi^{2}}\sqrt{z}(1+2z)
\left[{1\over z(1-z)}+{1\over z^{2}}\log(1-z)\right].
\label{(38)}
\end{eqnarray}
Note now that $\varphi_{c}^{(0)}(x)$ in Eq. (32), being a solution
of the homogeneous vector wave equation (11), admits the Huygens'
principle representation \cite{APNYA-9-220}
\begin{equation}
\varphi_{c}^{(0)}(x)=\int_{\Sigma'}\sqrt{-\gamma(x')}
\Bigr[G_{cb'}\varphi_{\; \; \; \; \; \; ;m'}^{(0)b'}
-G_{cb';m'}\varphi^{(0)b'}\Bigr]g^{m'l'} d\Sigma_{l'}^{'},
\label{(39)}
\end{equation}
where, from Sec. 3 and the present section,
\begin{equation}
G_{cb'}=\alpha g_{cb'}+\beta \mu_{;c} \mu_{;b'}={1\over 2}K_{cb'},
\label{(40)}
\end{equation}
\begin{equation}
G_{cb';m'}={1\over 2}K_{cb';m'}.
\label{(41)}
\end{equation}
Unlike the work in \cite{APNYA-9-220}, we here advocate the use of the Green
function (15) rather than the sum, over all distinct geodesics
between $x$ and $x'$, of the Hadamard functions.
To lowest order in the cosmological constant $\Lambda$, Eq. (39) may be
used to approximate the desired solution of Eq. (32) in the form
\begin{equation}
\varphi_{c}(x)=\varphi_{c}^{(0)}(x)
+\Lambda \int K_{c}^{\; \; a'} \varphi_{a'}^{(0)}
\sqrt{-\gamma(x')}d^{4}x'+{\rm O}(\Lambda^{2}).
\label{(42)}
\end{equation}
Omitting indices for simplicity, the general algorithm for solving Eq.
(32), here re-written in the form
\begin{equation}
\varphi=\varphi^{(0)}+\Lambda \int K \varphi,
\label{(43)}
\end{equation}
would be instead
\begin{equation}
\varphi_{1}=\varphi^{(0)}+\Lambda \int K \varphi^{(0)},
\label{(44)}
\end{equation}
\begin{equation}
\varphi_{2}=\varphi^{(0)}+\Lambda \int K \varphi_{1}
=\varphi^{(0)}+\Lambda \int K \varphi^{(0)}
+\Lambda^{2} \int \int K K \varphi^{(0)},
\label{(45)}
\end{equation}
\begin{equation}
\varphi_{n}=\varphi^{(0)}+\sum_{j=1}^{n}\Lambda^{j}
\int ... \int K^{j} \varphi^{(0)},
\label{(46)}
\end{equation}
\begin{equation}
\varphi= \lim_{n \to \infty} \varphi_{n}.
\label{(47)}
\end{equation}

\section{Separation of variables for the vector wave equations}

Consider now the de Sitter metric in standard spherical coordinates
\begin{equation}
ds^{2}=-fdt^{2}+\frac{1}{f}dr^{2}
+r^{2}(d\theta^{2}+\sin^{2}\theta d\phi^{2}),
\label{(48)}
\end{equation}
where $f \equiv 1-H^{2}r^{2}$ and $H$ is the Hubble 
constant. This metric satisfies the vacuum
Einstein equations with nonvanishing cosmological constant $\Lambda$
such that $H^{2}={\Lambda \over 3}$. Moreover, the timelike unit
normal vector fields $n$ to the $t={\rm constant}$ hypersurfaces
\begin{equation}
n={\partial \over \partial t}
\label{(49)}
\end{equation}
form a geodesic and irrotational congruence. The 3-metric induced on
the $t={\rm constant}$ hypersurfaces turns out to be conformally flat,
and the $(1,1)$ form of the spacetime Ricci tensor is simply given by
\begin{equation}
R_{a}^{\; b}=3H^{2}\delta_{a}^{\; b}.
\label{(50)}
\end{equation}
From the previous sections it is clear we are interested in the
generalized wave equation
\begin{equation}
-\cstok{\ }X_{a}+\epsilon R_{a}^{\; b}X_{b}
=\left(-\cstok{\ }+\epsilon {R\over 4}\right)X_{a}=0,
\label{(51)}
\end{equation}
where $\epsilon \equiv \pm 1$. By virtue of the
spherical symmetry of de Sitter
spacetime these equations should be conveniently wtitten by using the
expansion of $X$ in vector harmonics \cite{Lifs63, Gerl78, Jant78}.
Following Zerilli \cite{Zeri74} we have
\begin{eqnarray}
X&=& Y(\theta)e^{-i(\omega t - m \phi)}
[f_{0}(r)dt+f_{1}(r)dr] \nonumber \\
&+& e^{-i(\omega t - m \phi)}
\left[-{mr \over \sin \theta}f_{2}(r)Y(\theta)
+f_{3}(r){dY \over d\theta}\right]d\theta \nonumber \\
&+& i e^{-i(\omega t - m \phi)} \left[-r \sin \theta f_{2}(r)
{dY \over d\theta}+m f_{3}(r)Y(\theta)\right]d\phi,
\label{(52)}
\end{eqnarray}
with $Y(\theta)$ solution of the spherical harmonics equation
\begin{equation}
\left[{d^{2}\over d\theta^{2}}+\cot \theta {d \over d\theta}
+\left(-{m^{2}\over \sin^{2}\theta}+L \right)\right]Y=0,
\label{(53)}
\end{equation}
with $L\equiv l(l+1)$.
These equations lead to a system of coupled ordinary differential
equations for the functions $f_{0},f_{1},f_{3}$, besides a decoupled
equation for $f_{2}$ ($f_{2}$ being related to the transverse
part of $X$). Indeed, for $l\not=0,1$ (the latter case being trivial), we have
\begin{eqnarray}
\label{(54)}
{d^{2}f_{0}\over dr^{2}}&=&-{2\over r}{df_{0}\over dr}
-{1\over r^{2}f^{2}}\Bigr[\omega^{2}r^{2}-Lf+3f(1-f)(1-\epsilon)\Bigr]
f_{0}+{2i \omega(f-1)\over rf}f_{1}, \\
{d^{2}f_{1}\over dr^{2}}&=& {2(2-3f)\over rf}{df_{1}\over dr}
-{1\over r^{2}f^{2}}\Bigr[\omega^{2}r^{2}-Lf
-3\epsilon f(1-f)+f(f-3)\Bigr]f_{1} \nonumber \\
&-& {2i \omega (1-f)\over rf^{3}}f_{0}
-{2L \over r^{3}f}f_{3},\\
\label{(55)}
{d^{2}f_{2}\over dr^{2}}&=&{2(1-2f)\over rf}{df_{2}\over dr}
-{1\over r^{2}f^{2}}\Bigr[\omega^{2}r^{2}-Lf+f(1-f)(1-3\epsilon)\Bigr]
f_{2},
\label{(56)}\\
{d^{2}f_{3}\over dr^{2}}&=&{2(1-f)\over rf}{df_{3}\over dr}
-{1\over r^{2}f^{2}}\Bigr[\omega^{2}r^{2}-Lf+3f(1-f)(1-\epsilon)
\Bigr]f_{3}-{2\over r}f_{1}.
\label{(57)}
\end{eqnarray}
Equation (56) for $f_{2}(r)$ can be easily integrated in terms of
hypergeometric functions. In fact, assuming
\begin{equation}
f_{2}(r)=f^{-i \omega/(2H)}\psi(r), 
\label{(58)}
\end{equation}
the resulting equation for $\psi$ reads as
\begin{equation}
{d^{2}\psi \over dr^{2}}=-{2i \over rf}\Bigr(2iH^{2}r^{2}
-i+\omega Hr^{2}\Bigr){d\psi \over dr}
-{1\over r^{2}f^{2}}\Bigr[\omega^{2}r^{2}-L
+(1-3 \epsilon)H^{2}r^{2}+3i \omega Hr^{2} \Bigr],
\label{(59)}
\end{equation}
the solution of which is, in general, of the form
\begin{equation}
\psi(r)=C_{1}r^{l}{ }_{2}F_{1} \left(a_{-},a_{+},
{3\over 2}+l,H^{2}r^{2} \right)
+C_{2}r^{-1-l}{ }_{2}F_{1} \left(a_{+},a_{-},
{1\over 2}- l,H^{2}r^{2} \right),
\label{(60)}
\end{equation}
where we have defined
\begin{equation}
a_{\pm} \equiv -{1\over 4}\biggr[{2i \omega \over H}
-3 -2l \pm (13-12 \epsilon)^{1/2} \biggr].
\label{(61)}
\end{equation}
Thus, when $\epsilon=1$, which corresponds to studying the vector wave
equation (7), one finds
\begin{equation}
a_{\pm}=-{1\over 4} \left({2i \omega \over H}-3-2l \pm 1 \right),
\label{(62)}
\end{equation}
whereas on taking $\epsilon=-1$, i.e. our consistency Eq. (8), one gets
\begin{equation}
a_{\pm}=-{1\over 4}\left({2i \omega \over H}-3 -2l \pm 5 \right).
\label{(63)}
\end{equation}
The Lorenz gauge condition $\nabla_{\alpha}X^{\alpha}=0$, which only
supplements Eq. (7), reduces instead to
\begin{equation}
f_{3}={r^{2}f \over L}{df_{1}\over dr}
-{2r(1-2f)\over L}f_{1}+{\omega r^{2}i \over Lf}f_{0}.
\label{(64)}
\end{equation}
The latter condition can be used, in principle, to obtain closed-form
solutions of the various $f_{0}(r),f_{1}(r),f_{3}(r)$.

\section{Wave equation for metric perturbations}

Although the vector wave equations in de Sitter spacetime are already
considerably involved, the final step consists in
studying the invertible wave operator $P_{ab}^{\; \; \; cd}$ on metric
perturbations. On considering the DeWitt supermetric
\begin{equation}
E^{abcd} \equiv \gamma^{a(c} \; \gamma^{d)b}
-{1\over 2}\gamma^{ab}\gamma^{cd},
\label{(65)}
\end{equation}
the de Donder gauge in Eq. (4) can be re-expressed in the form
\begin{equation}
\Phi_{a}(h)=E_{a}^{\; bcd}\nabla_{b}h_{cd},
\label{(66)}
\end{equation}
and the resulting Lichnerowicz operator
\cite{BSMFA-92-11}, \cite{NUPHA-B146-90}
on metric perturbations, obtained by
expansion of the Einstein--Hilbert action to quadratic order in
$h_{ab}$, subject to $\Phi_{a}(h)=0$, reads as \cite{Moss1996}
\begin{equation}
P_{ab}^{\; \; \; cd} \equiv E_{ab}^{\; \; \; cd}(-\cstok{\ }+R)
-2E_{ab}^{\; \; \; lf}R_{\; lhf}^{c}\gamma^{dh}
-E_{ab}^{\; \; \; ld}R_{l}^{\; c}-E_{ab}^{\; \; \; cl}R_{l}^{\; d}.
\label{(67)}
\end{equation}

A wave equation for metric perturbations is therefore given by
(see the Introduction)
\begin{equation}
P_{ab}{}^{cd}h_{cd}=0\ ,
\label{(68)}
\end{equation}
where \cite{Moss1996}
\begin{eqnarray}
P_{ab}{}^{cd}&=&E_{ab}{}^{cd}(-\cstok{\ } +R)
-2E_{ab}{}^{lf}R^c{}_{lhf}\gamma^{dh}-E_{ab}{}^{ld}R_l{}^c
-E_{ab}{}^{cl}R_l{}^d\nonumber \\
&=& E_{ab}{}^{cd}(-\cstok{\ } +R)-2E_{ab}{}^{lf}
R_{ml}{}^{dn}\gamma^{mc}\gamma_{nf}-\frac{R}{4}E_{ab}{}^{cd}
-\frac{R}{4}E_{ab}{}^{cd}\nonumber \\
&=& E_{ab}{}^{cd}(-\cstok{\ } +R)-\frac{R}{6}E_{ab}{}^{lf}
\delta_{ml}^{dn}\gamma^{mc}\gamma_{nf}-\frac{R}{2}E_{ab}{}^{cd}\nonumber \\
&=& E_{ab}{}^{cd}\left(-\cstok{\ } +\frac12 R\right)
-\frac{R}{6}E_{ab}{}^{lf}\delta_{ml}^{dn}\gamma^{mc}\gamma_{nf}\nonumber \\
&=& E_{ab}{}^{cd}\left(-\cstok{\ } +\frac12 R\right)+\frac{R}{6}E_{ab}{}^{cd}
+\frac{R}{6}\gamma_{ab}\gamma^{cd}\nonumber \\
&=& E_{ab}{}^{cd}\left(-\cstok{\ } +\frac23 R\right)
+\frac{R}{6}\gamma_{ab}\gamma^{cd}.
\label{(69)}
\end{eqnarray}
The wave equation then becomes
\begin{equation}
0=P_{ab}{}^{cd}h_{cd}=\left(-\cstok{\ } +\frac23 R\right)
\bar h_{ab}+\frac{R}{6}\gamma_{ab}h,
\label{(70)}
\end{equation}
or
\begin{equation}
\left(-\cstok{\ } +\frac23 R\right)\bar h_{ab}
-\frac{R}{6}\gamma_{ab}{\bar h} =0 ,
\label{(71)}
\end{equation}
implying also
\begin{equation}
\left(-\cstok{\ } +\frac23 R\right)\bar h-\frac{2}{3}R{\bar h}=0,
\label{(72)}
\end{equation}
that is
\begin{equation}
\cstok{\ } {\bar h}=0 ,
\label{(73)}
\end{equation}
after contraction with $\gamma^{ab}$.

\subsection{Even metric perturbations}

Metric perturbations of even parity can be written in the form
\begin{eqnarray}
h_{00}&=& f e^{-i(\omega t -m\phi)}H_0(r)Y(\theta),\nonumber \\
h_{01}&=& e^{-i(\omega t -m\phi)}H_1(r)Y(\theta),\nonumber \\
h_{02}&=& e^{-i(\omega t -m\phi)}h_0(r)\frac{dY}{d\theta},\nonumber \\
h_{03}&=& im e^{-i(\omega t -m\phi)}h_0(r)Y(\theta),\nonumber \\
h_{11}&=& \frac{1}{f} e^{-i(\omega t -m\phi)}H_2(r)Y(\theta),\nonumber \\
h_{12}&=& e^{-i(\omega t -m\phi)}h_1(r)\frac{dY}{d\theta},\nonumber \\
h_{13}&=& im e^{-i(\omega t -m\phi)}h_1(r)Y(\theta),\nonumber \\
h_{22}&=& r^2 e^{-i(\omega t -m\phi)}\left[K(r)Y(\theta)
+G(r)\frac{d^2Y}{d\theta^2}\right],\nonumber \\
h_{23}&=& im r^2 G(r) e^{-i(\omega t -m\phi)}
\left[\frac{dY}{d\theta}-\cot \theta Y(\theta)\right],\nonumber \\
h_{33}&=& r^2 e^{-i(\omega t -m\phi)}\left\{K(r)
\sin^2\theta Y(\theta)\right. \nonumber \\
&& \left. +G(r)\left[-m^2 Y(\theta)+\sin \theta
\cos \theta\frac{dY}{d\theta}\right]\right\} .
\label{(74)}
\end{eqnarray}
The wave equation (68)
leads to the following system of coupled differential equations:
\begin{eqnarray}
\frac{d^2H_0}{dr^2}&=&\frac{2}{rf}(2-3f)\frac{dH_0}{dr}
-\frac{1}{r^2f^2}[\omega^2r^2 +2(1-f)(1-4f)-fL]H_0\nonumber \\
&& +\frac{2H^2r}{f}\frac{dH_2}{dr}-\frac{4H^2r}{f}
\frac{dK}{dr}+\frac{2H^2rL}{f}\frac{dG}{dr}\nonumber \\
&& +\frac{2H^2L}{f}G-\frac{4H^2L}{rf}h_1-
\frac{2H^{2}(1-6f)}{f^{2}}H_2-\frac{4H^2}{f}K , \nonumber \\
\frac{d^2H_1}{dr^2}&=& \frac{2}{rf}(1-2f)
\frac{dH_1}{dr}-\frac{1}{r^2f^2}[\omega^2r^2-Lf-2(2-f^2)]H_1 \nonumber \\
&& -\frac{2L}{fr^3}h_0 -\frac{2i\omega rH^2}{f^2}(H_2+H_0), \nonumber \\
\frac{d^2H_2}{dr^2}&=&\frac{2(2-3f)}{rf}\frac{dH_2}{dr}
-\frac{1}{r^2f^2}[\omega^2r^2-Lf-10+6f+12(1-f)^2]H_2\nonumber \\
&& +\frac{2H^2r}{f}\frac{dH_0}{dr}-\frac{4H^2r}{f}
\frac{dK}{dr}+\frac{2rLH^2}{f}\frac{dG}{dr} \nonumber \\
&& +\frac{2L(3-2f)}{r^2f}G-\frac{4L}{r^3f}h_1
-\frac{4(3-2f)}{r^2f}K-\frac{2H^2(1-2f)}{f^2}H_0,\nonumber \\
\frac{d^2h_0}{dr^2}&=& -\frac{1}{r^2f^2}[\omega^2r^2-Lf-4f(1-f)]h_0
-\frac{2i\omega rH^2}{f}h_1-\frac{2}{r}H_1, \nonumber \\
\frac{d^2h_1}{dr^2}&=&\frac{6rH^2}{f}\frac{dh_1}{dr}
-\frac{1}{r^2f}[\omega^2r^2-Lf+10(1-f)^{2}-6+2f]h_{1}
+\frac{1}{rf^2}[2-(1+f)L]G\nonumber \\
&& -\frac{1+f}{rf^2}H_2+\frac{2}{rf}K+\frac{H^2r}{f^2}H_0, \nonumber \\
\frac{d^2G}{dr^2}&=& \frac{2}{rf}(1-2f)\frac{dG}{dr}
-\frac{1}{r^2f^2}[\omega^2r^2-Lf+2f(3f-2)]G
-\frac{4h_1}{r^3}, \nonumber \\
\frac{d^2K}{dr^2}&=&\frac{2(1-2f)}{rf}\frac{dK}{dr}
-\frac{1}{r^2f^2}[\omega^2r^2-Lf-2f(2-f)]K-\frac{2L}{r^2}G\nonumber \\
&& -\frac{2}{r^2f}H_2+\frac{2H^2}{f}H_0. \nonumber \\
\label{(75)}
\end{eqnarray}
To these equations one should add the de Donder gauge components
\begin{eqnarray}
\frac{dH_1}{dr}&=& \frac{i\omega L}{2f}G+\frac{L}{r^2f}h_0
-\frac{i\omega}{2f}H_2-\frac{i\omega}{f}\left(K+{H_{0}\over 2}\right)
+\frac{2(1-2f)}{rf}H_1, \nonumber \\
\frac{dH_2}{dr}&=& \frac{2(1-3f)}{rf}H_2-\frac{dH_0}{dr}
+2\frac{dK}{dr}-L\frac{dG}{dr}-\frac{2L}{r}G\nonumber \\
&& +\frac{2L}{r^2}h_1+\frac{4}{r}K+\frac{2H^2r}{f}H_0
-\frac{2i\omega}{f}H_1, \nonumber \\
\frac{dh_1}{dr}&=&\frac{2(1-2f)}{rf}h_1+\frac{L-2}{2f}G
-\frac{i\omega}{f^2}h_0+\frac{1}{2f}(H_2-H_0).
\label{(76)}
\end{eqnarray}
Metric perturbations of odd parity are instead found to vanish
identically.

\section{Concluding remarks}

A consistency analysis for gravitational waves in curved
background is not, by itself, new in the literature and it has
been outlined, for example, in Sec. II and Appendix B of Ref.
\cite{PHRVA-D66-064024}. However, that paper was mainly
concerned with gravitational instability in higher dimensions. In
this paper, we have endeavoured to provide explicit solution
formulae for the covector which solves the residual gauge problem
expressed by Eq. (6), and this has been accomplished in Secs. 4
and 5. Section 6 has then moved on to work out all equations obeyed
by metric perturbations subject to the de Donder gauge in the de
Sitter spacetime. Although the system (74)--(76) obeyed by metric
perturbations looks very complicated, our approach is suitable for
computer-assisted investigation of such equations, with the
possibility of investigating gravitational waves in inflationary
cosmology \cite{Hawk00} and/or other curved backgrounds relevant
for strong-gravity regimes. In particular, as discussed in
\cite{garcia}, a significant fraction of energy, in the form of a
stochastic background of  gravitational waves could emerge during
the reheating after inflation. In this situation, the exact
classification of gravitational wave solutions (as in Sec. 6)
could be crucial in order to discriminate among the various
signals. In a forthcoming paper, starting from the presented
solutions, we will discuss the problem of generation
and production of gravitational waves in the 
de Sitter background \cite{felix}.

\appendix
\section{Bivectors and biscalars}

In Eq. (13), $g_{\; c'}^{a}$ is the geodesic parallel displacement
bivector (in general, bitensors behave as a tensor both at $x$ and
at $x'$) which effects parallel displacement of vectors along the
geodesic from $x'$ to $x$. In general, it is defined by the
differential equations
\begin{equation}
\sigma^{;b} \; g_{\; c';b}^{a}=\sigma^{;b'} \;
g_{\; c';b'}^{a}=0,
\label{(A1)}
\end{equation}
jointly with the coincidence limit
\begin{equation}
\lim_{x' \to x}g_{\; c'}^{a}=\Bigr[g_{\; c'}^{a} \Bigr]
=\delta_{\; c}^{a}.
\label{(A2)}
\end{equation}
The bivector $g_{\; c'}^{a}$, when acting on a vector $B^{c'}$ at
$x'$, gives therefore the vector ${\overline B}^{a}$ which is obtained by
parallel transport of $B^{c'}$ to $x$ along the geodesic connecting
$x$ and $x'$, i.e.
\begin{equation}
{\overline B}^{a}=g_{\; c'}^{a} \; B^{c'}.
\label{(A3)}
\end{equation}

In Eq. (14), $\bigtriangleup(x,x')$ is a biscalar built from the
Van Vleck--Morette determinant
\begin{equation}
D(x,x') \equiv {\rm det} (\sigma_{; ab'})
\label{(A4)}
\end{equation}
according to
\begin{equation}
\bigtriangleup(x,x') \equiv {1\over \sqrt{-\gamma(x)}} D(x,x')
{1\over \sqrt{-\gamma(x')}}.
\label{(A5)}
\end{equation}
The biscalar $\bigtriangleup(x,x')$ has unit coincidence limit:
$[\bigtriangleup]=1$; as a function of $x$ (resp. $x'$), it becomes
infinite on any caustic formed by the geodesics emanating from $x'$
(resp. $x$). When $\bigtriangleup$ diverges in this way, $x$ and $x'$
are said to be conjugate points \cite{DeWi84}.

\acknowledgments

G. Esposito is grateful to the Dipartimento di Scienze Fisiche of
Federico II University, Naples, for hospitality and support,
and to CNR for financial support.

\end{document}